\documentclass[a4paper]{article}

\usepackage{amssymb}
\usepackage{graphicx}
\usepackage{subfigure}
\usepackage{epsfig}
\usepackage{floatflt}
\usepackage{afterpage}

\newcommand{\epm}{\ensuremath{e^{+}e^{-}}}
\newcommand{\pne}{\ensuremath{\pi^{+}\pi^{-}2\pi^{0}}}
\newcommand{\pch}{\ensuremath{2\pi^{+}2\pi^{-}}}
\newcommand{\ompi}{\ensuremath{\omega\pi^{0}}}
\newcommand{\api}{\ensuremath{a_{1}\pi} }
\newcommand{\anpi}{\ensuremath{a_{1}(1260)\pi} }
\newcommand{\sigrho}{\ensuremath{\rho \sigma} }
\newcommand{\hpi}{\ensuremath{h_{1}\pi} }
\newcommand{\hnpi}{\ensuremath{h_{1}(1170)\pi} }
\newcommand{\aapi}{\ensuremath{a_{2}\pi} }
\newcommand{\aanpi}{\ensuremath{a_{2}(1320)\pi} }
\newcommand{\piprimpi}{\ensuremath{\pi^{\prime} \pi} }
\newcommand{\pinpi}{\ensuremath{\pi(1300)\pi} }
\newcommand{\rhorho}{\ensuremath{\rho ^{+}\rho ^{-}}}

\newcommand{\mel}{|\mathfrak{M}(\vec{P}_{i},\vec{A})|^{2}}
\newcommand{\jel}[1]{\mathbf{J}^{+-00}_{#1 }}
\newcommand{\wfun}{w(\vec{P}_{i},\vec{A})}

\newcommand{\beq}{\begin{equation}}
\newcommand{\eeq}{\end{equation}}
\newcommand{\eq}[1]{(\ref{#1})}

\newcommand{\beqn}{\begin{eqnarray}}
\newcommand{\eeqn}{\end{eqnarray}}
\newcommand{\dst}{&\displaystyle}
\newcommand{\eps}{\mbox{$\varepsilon$}}
\newcommand{\T}{{\bf t}}

\newcommand{\p}{\mbox{${\bf p}$}}

\begin{document}

\newpage
\thispagestyle{empty}

\begin{center}
{\large Siberian Branch of Russian Academy of Sciences\\[2mm]
{\normalsize  BUDKER INSTITUTE OF NUCLEAR PHYSICS}
\\[15mm]
R.R.~Akhmetshin, E.V.~Anashkin, M.~Arpagaus, V.M.~Aulchenko, 
V.Sh.~Banzarov, L.M.~Barkov,
N.S.~Bashtovoy, A.E.~Bondar, D.V.~Bondarev, A.V.~Bragin,
D.V.~Chernyak, A.G.~Chertovskikh, A.S.~Dvoretsky,
S.I.~Eidelman, G.V.~Fedotovitch, 
N.I.~Gabyshev,
A.A.~Grebeniuk, D.N.~Grigoriev, V.W.~Hughes, P.M.~Ivanov, S.V.~Karpov,
 V.F.~Kazanin, 
B.I.~Khazin, I.A.~Koop, M.S.~Korostelev, P.P.~Krokovny, L.M.~Kurdadze, 
A.S.~Kuzmin, 
I.B.~Logashenko, P.A.~Lukin, A.P.~Lysenko, 
K.Yu.~Mikhailov, A.I.~Milstein,
 I.N.~Nesterenko,
V.S.~Okhapkin, A.V.~Otboev, E.A.~Perevedentsev, A.A.~Polunin, A.S.~Popov,
T.A.~Purlatz, N.I.~Root, A.A.~Ruban, N.M.~Ryskulov, A.G.~Shamov, 
Yu.M.~Shatunov, A.I.~Shekhtman,
B.A.~Shwartz, A.L.~Sibidanov,
V.A.~Sidorov, A.N.~Skrinsky, V.P.~Smakhtin, I.G.~Snopkov,
E.P.~Solodov, P.Yu.~Stepanov, A.I.~Sukhanov, J.A.~Thompson, V.M.~Titov, 
A.A.~Valishev, Yu.V.~Yudin, S.G.~Zverev

\vskip 15mm
\Large{ \boldmath \bf
$a_{1}(1260) \pi$  dominance in the process 
$\epm \to 4\pi$ at energies 1.05--1.38 GeV}
\\[15mm]
\vfill
Budker INP 98-83\\
Novosibirsk\\[1mm]
1998 }
\end{center}

\newpage
\thispagestyle{empty}
\vfill
\begin{center}
\Large{ \boldmath \bf
$a_{1}(1260) \pi$  dominance in the process 
$\epm \to 4\pi$ at energies 1.05--1.38 GeV}
\\[15mm]
\end{center}
\begin{center}
R.R.~Akhmetshin, E.V.~Anashkin, M.~Arpagaus, V.M.~Aulchenko, 
V.Sh.~Banzarov, L.M.~Barkov,
N.S.~Bashtovoy, A.E.~Bondar, D.V.~Bondarev, A.V.~Bragin,
D.V.~Chernyak, A.G.~Chertovskikh, A.S.~Dvoretsky,
S.I.~Eidelman, G.V.~Fedotovitch, 
N.I.~Gabyshev,
A.A.~Grebeniuk, D.N.~Grigoriev, P.M.~Ivanov, S.V.~Karpov,
 V.F.~Kazanin, 
B.I.~Khazin, I.A.~Koop, M.S.~Korostelev, P.P.~Krokovny, L.M.~Kurdadze, 
A.S.~Kuzmin, 
I.B.~Logashenko, P.A.~Lukin, A.P.~Lysenko, 
K.Yu.~Mikhailov, A.I.~Milstein,
 I.N.~Nesterenko,
V.S.~Okhapkin, A.V.~Otboev, E.A.~Perevedentsev, A.A.~Polunin, A.S.~Popov,
T.A.~Purlatz, N.I.~Root, A.A.~Ruban, N.M.~Ryskulov, A.G.~Shamov, 
Yu.M.~Shatunov, A.I.~Shekhtman,
B.A.~Shwartz, A.L.~Sibidanov,
V.A.~Sidorov, A.N.~Skrinsky, V.P.~Smakhtin, I.G.~Snopkov,
E.P.~Solodov, P.Yu.~Stepanov, A.I.~Sukhanov, V.M.~Titov, 
A.A.~Valishev, Yu.V.~Yudin, S.G.~Zverev
\\
  Budker Institute of Nuclear Physics, Novosibirsk, 630090, Russia
\end{center}
\begin{center}
                J.A.Thompson
\\
                University of Pittsburgh, Pittsburgh, PA 15260, USA
\end{center}
\begin{center}
                V.W.Hughes
\\
                    Yale University, New Haven, CT 06511, USA
\end{center}
\vfill
\begin{abstract}

   First results of the study of the process $\epm \to 4\pi$
by the CMD-2 collaboration at VEPP-2M are presented for the energy range 
1.05--1.38 GeV. Using an integrated luminosity of 5.8 $pb^{-1}$, energy
dependence of the processes $\epm \to \pne$ and $\epm \to \pch$ 
has been measured.
Analysis of the differential distributions demonstrates the
dominance of the $a_1 \pi$ and $\omega \pi$ intermediate states.
Upper limits for the contributions of other alternative mechanisms
are also placed.

\end{abstract}

\vfill
\newpage

\section{\bf Introduction. }

Investigation of $\epm$ annihilation into hadrons at low energies
provides unique information about interaction of light quarks
and spectroscopy of their bound states. At present the energy
range below $J/\psi$ can not be satisfactorily described by QCD.
Future progress in our understanding of the phenomena in this
energy range is impossible without accumulation of
experimental data vitally important to check the predictions of existing
theoretical models. In addition, the total cross section
of $\epm$ annihilation into hadrons at low energies as well as the 
cross sections of exclusive channels are necessary
for precise calculations of various effects. These include strong interaction 
contributions to vacuum polarization for (g-2)$_{\mu}$ and 
$\alpha(M_{Z}^{2})$ \cite{ej}, tests of standard model by the hypothesis 
of conserved vector current (CVC) relating $\epm \to$ hadrons to 
hadronic $\tau$-lepton decays~\cite{CVC1,CVC2},
determination of the QCD parameters based on QCD sum rules~\cite{SVZ} etc. 

The energy behavior of the total cross section as well as that of the
cross sections for exclusive channels is complicated and characteristic of 
various broad overlapping resonances 
(e.g. $\rho, \omega, \phi$ recurrencies) 
with numerous common decay channels having energy thresholds just
in this energy range. Presence of the broad resonances in the
intermediate state necessitates consideration of the quasistationary
states and makes effects of their interference important. 

Until
recently the investigation of $\epm$ annihilation into hadrons
was restricted by measurements of the cross sections only. Appearance
of the new detectors with a large solid angle operating at high luminosity 
colliders and providing very large data samples opens qualitatively new 
possibilities for the investigation of the multihadronic production
in $\epm$ annihilation.

Production of four pions is one of the dominant processes of
$\epm$ annihilation into hadrons in the energy range from 1.05 to 2.5 GeV. 
For the first time it was 
observed in Frascati~\cite{fr1} and Novosibirsk~\cite{nov1}.
First experiments with limited data samples allowed one
to qualitatively study the new phenomenon of multiple production 
of hadrons and estimate the magnitude of the corresponding cross sections.
Subsequent measurements by different groups in Frascati, Orsay and
Novosibirsk (see the references in~\cite{ej})
provided more detailed information on the energy dependence of the 
cross sections of the processes $\epm \to \pch$ and $\epm \to \pne$
in comparison with the previous measurements.  

One of the main difficulties in the experimental studies of
four pion production was caused by the existence of
different intermediate states via which the final state could be
produced, such as
\beqn
\epm & \to & \omega \pi \label{eq1}\\
\epm & \to & \rho \sigma \\
\epm & \to & a_{1}(1260) \pi \\ 
\epm & \to & h_{1}(1170) \pi \\
\epm & \to & \rho^{+} \rho^{-} \\
\epm & \to & a_{2}(1320) \pi \\
\epm & \to & \pi(1300) \pi \label{eq7}
\eeqn

The relative contributions of the above mentioned processes can't
be obtained without the detailed analysis of the process dynamics. 
First attempts of this type were performed by MEA~\cite{mea}
and DM1~\cite{dm1} in the energy range above 1.4 GeV and
OLYA~\cite{olya2} and CMD~\cite{cmd} below 1.4 GeV and it was shown that 
the $\pch$ final state is dominated by the $\rho^{0} \pi^+ \pi^-$ 
mechanism. ND~\cite{opi} measured the cross section of  
$\omega \pi$ production from 1.0 to 1.4 GeV with a magnitude
which was confirmed by 
the subsequent $\tau$ decay studies at 
\nolinebreak{ARGUS~\cite{argus}} as well as by
more recent results from CLEO~\cite{cleo} and ALEPH~\cite{aleph}.  

Later the DM2 group tried to perform partial wave analysis (PWA) of the 
mode with four charged pions \cite{dm2} in the energy range 1.35 to 
2.40 GeV. Their analysis was based
on the momentum distributions only while the angular dependence as well as
interference between different waves were not taken into account. 
Although they obtained some evidence for the presence of
$a_1(1260) \pi$ and $\rho \sigma$ states, a mechanism for a substantial 
part of the cross section was not determined. PWA for the mode
$\pne$ was not performed because of the insufficient number
of completely reconstructed events.  

The abundance of various possible mechanisms and their 
complicated interference results in the necessity of simultaneous
analysis of two possible final states ($\pch$ and $\pne$) which requires
a general purpose detector capable of measuring
energies and angles of both charged and neutral particles.
The first detector of this type operating in the energy
range below 1.4 GeV is 
the CMD-2 detector at VEPP-2M collider in Novosibirsk \cite{cmddec}.
In this work, we present results from a model-dependent analysis
of both possible channels
in $\epm$ annihilation into four pions based on data collected with 
the CMD-2 detector.
To describe 
four pion production we used a simple model assuming 
quasitwoparticle intermediate states and taking into account 
the important effects of the identity
of the final pions 
as well as the interference of all 
possible amplitudes.

\section{\boldmath Data sample and event selection}

The analysis described here is based on 5.8 pb$^{-1}$ of $\epm$
data collected at center-of-mass energies 2$E_{beam}$ from 1.05 up to
1.38 GeV. The data were recorded at the   
VEPP-2M $\epm$ collider of the Budker Institute of Nuclear
Physics in Novosibirsk, Russia with the CMD-2 detector in 1997.
The energy range mentioned above was 
scanned twice with a step of 20 MeV: first by increasing energy from 
1.05 to 1.37 GeV and then by decreasing energy from 1.38 to 1.06 GeV.

The CMD-2 is a general purpose detector 
consisting of a   drift  chamber (DC)
with about 250 $\mu$ resolution transverse to the beam and
proportional Z-chamber used for trigger, both inside a thin (0.4 $X_0$)
superconducting solenoid with a field of 1 T. Photons are detected in 
the barrel CsI calorimeter with 8-10 $\%$ energy resolution
and the endcap BGO calorimeter with 6 $\%$ energy resolution.
More details on the detector can be found elsewhere \cite{cmddec}.

\subsection{Event selection}

The present analysis is based on completely reconstructed $\pne$
events. At the initial stage, events with one primary vertex
with two opposite sign tracks and four or more reconstructed photons
were selected. Both tracks should come from the interaction region:
a distance from the track trajectories to the beam axis should be less than
0.3 cm and the vertex position along the beam axis should be inside
$\pm$ 10 cm. To reject background from collinear events,
the acollinearity angle between tracks in the $R-\phi$ plane
should be greater than 0.1 radians. Both tracks are required in the
DC fiducial volume:  a $\theta$ angle should be inside $0.54\div\pi-0.54$
radians. Clusters of energy deposition that are not matched
with charged track projection are paired to form $\pi^0$ candidates.
These showers must have energies greater than 20 MeV, and 
invariant mass of the photon pair must lie within $3\sigma$ of
the $\pi^0$ mass where 
$\sigma$ varies between $(5 \div 10)$ MeV.
After that a kinematic fit was
performed assuming the $\pne$ hypothesis for all possible $\pi^0$
pairs.
For further analysis the combination with $min(\chi^{2})$ was
selected under the condition $\chi^{2}/ndf~<~2.5$. 
After such selection 
22128 events remained in the energy range under study.

\subsection{Final event sample}

\begin{figure}[htbp]
\centering
\epsfig{figure=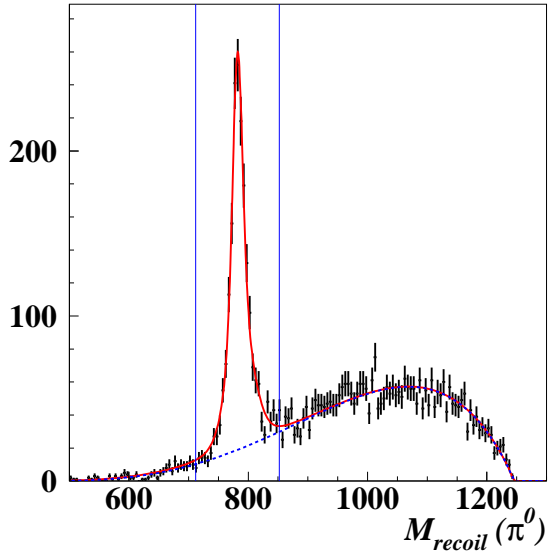,width=0.7\linewidth}
\caption{
Distribution over the $\pi^{0}$ recoil mass
for $\pne$
}
\label{omega}
\end{figure}

To understand the dynamics of the process we studied the distribution over 
the recoil mass for one of the $\pi^0$'s. 
Figure~\ref{omega} shows this distribution 
at the beam energy of 690 MeV. Each event gives two entries to the
histogram corresponding to two $\pi^{0}$. 
A signal from the $\ompi$ final state is clearly seen. 
Points with errors are the data while the smooth line is our fit
corresponding to the sum of a Breit-Wigner $\omega$ signal convoluted 
with detector resolution and
a smooth combinatorial background.
The detector contribution
to the signal width is about 10 MeV.
The number of events under the $\omega$ peak accounts for only $\sim 60\%$ 
of the observed events that indicates at the existence of additional 
intermediate states.

For further analysis the data sample was subdivided into
two classes:
\begin{enumerate}
\renewcommand{\theenumi}{\Roman{enumi}}
\item $min(|M_{recoil}(\pi^{0})-M_{\omega}|)~<~70~ MeV$ \label{class1}
\item $min(|M_{recoil}(\pi^{0})-M_{\omega}|)~>~70~ MeV$ \label{class2}
\end{enumerate}
where $M_{\omega}$ is the $\omega$ mass.
The first class contains mostly $\ompi$ events while their
admixture in the second class
is relatively small, about $(1\div5)\%$ 
depending on the beam energy.

\begin{figure}[htbp]
\epsfig{figure=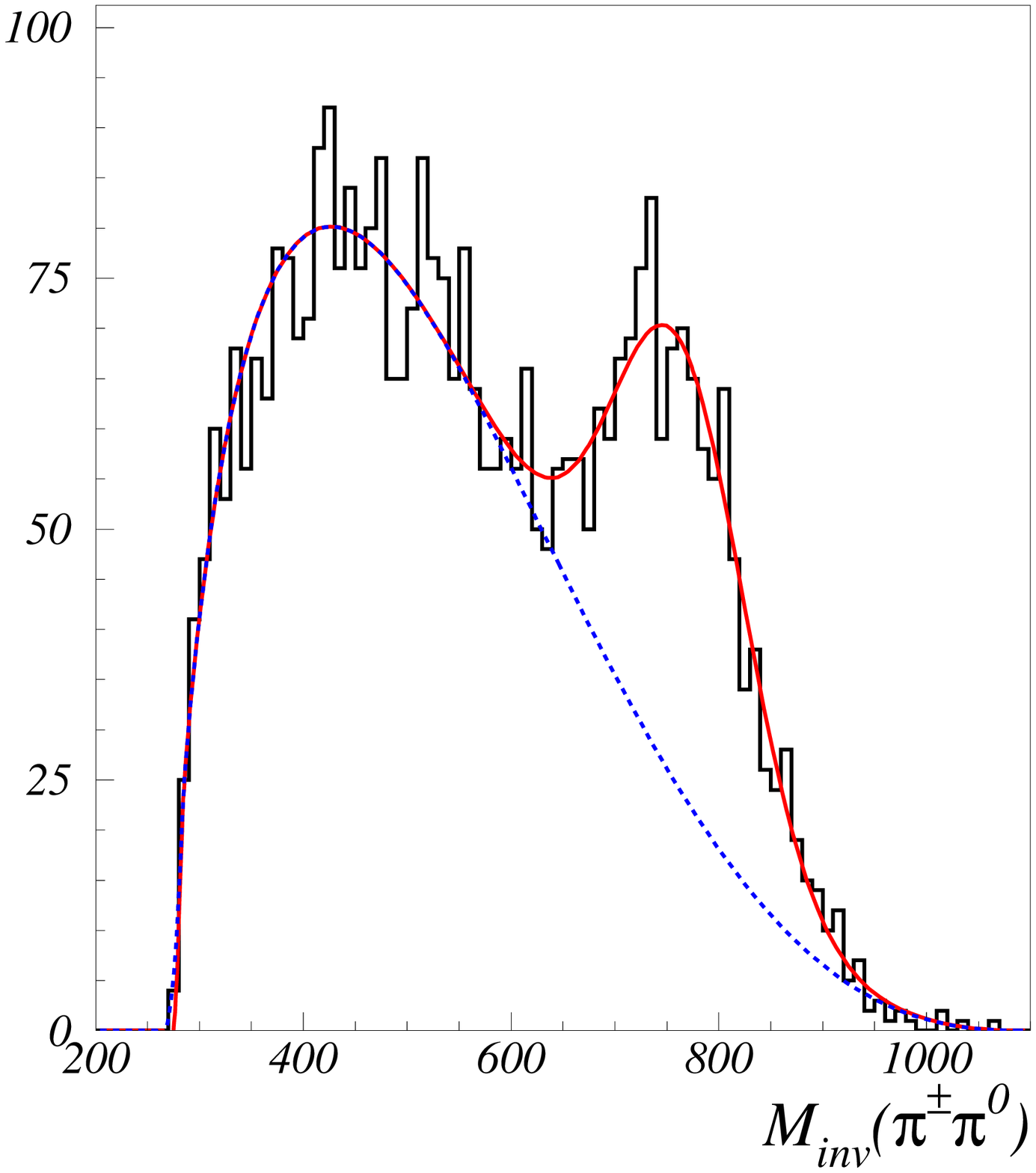,width=0.5\textwidth}
\hfill
\epsfig{figure=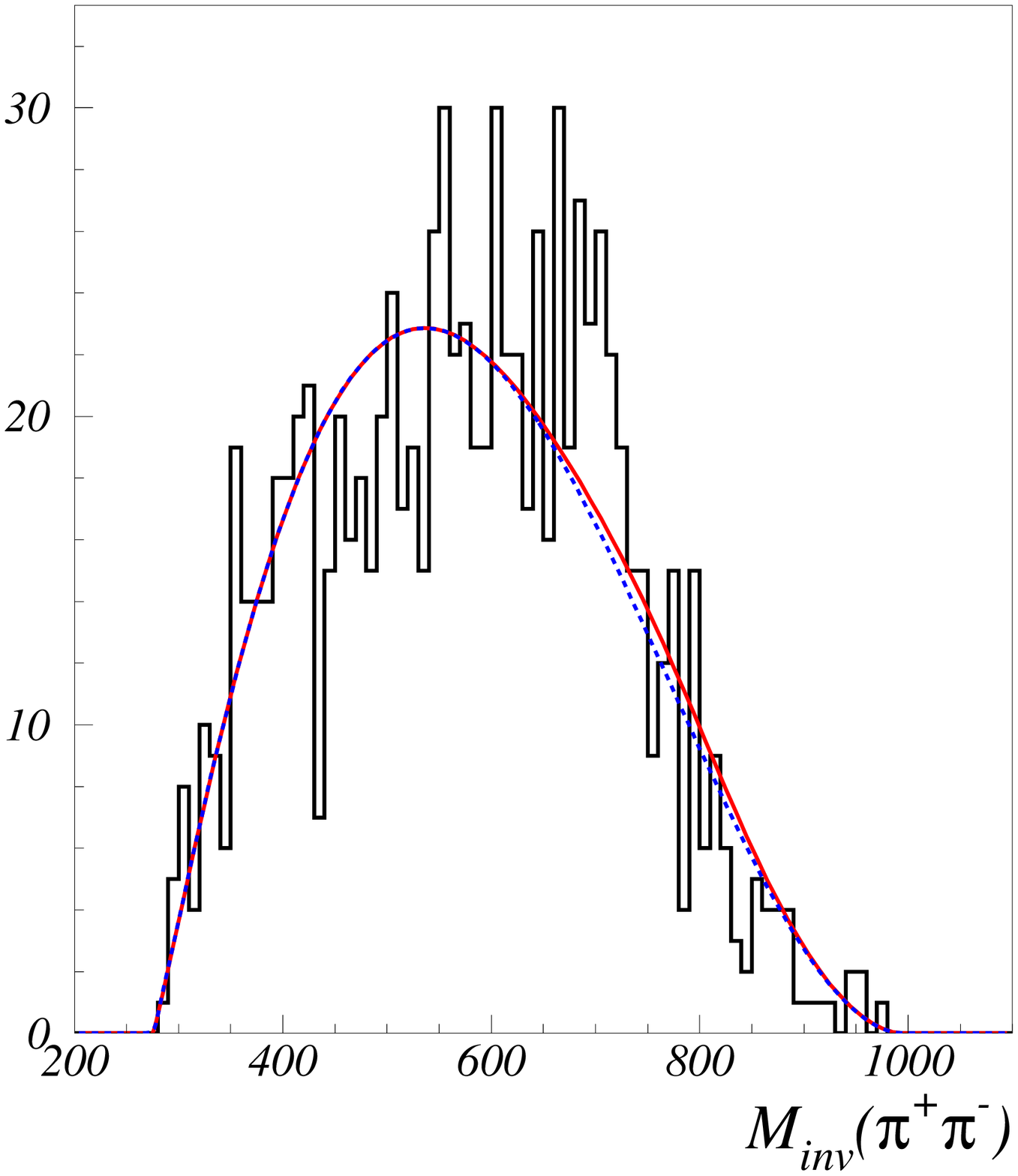,width=0.5\textwidth}
\\
\parbox[t]{0.47\textwidth}
{
\caption{
Distribution over $M_{inv}(\pi^{\pm}\pi^0)$ 
for $\pne$ events from the class (\ref{class2})
}
\label{rhopm}
}
\hfill
\hfill
\parbox[t]{0.47\textwidth}
{
\caption{
Distribution over $M_{inv}(\pi^+\pi^-)$ 
for $\pne$ events from the class (\ref{class2})
}
\label{rhozero}
}
\\
\end{figure}

Figs.~\ref{rhopm} and \ref{rhozero} show distributions over
$M_{inv}(\pi^{\pm}\pi^{0})$ and
$M_{inv}(\pi^{+}\pi^{-})$ for the events in the second class.
In the spectrum of  $M_{inv}(\pi^{\pm}\pi^{0})$ one can see
a clear signal of $\rho^{\pm}$ while that for 
 $M_{inv}(\pi^{+}\pi^{-})$
is relatively smooth and no signal from the $\rho^0$ is observed.
The solid lines show our fit including 
smooth combinatorial background and gaussian $\rho$ ~signals.
Presence of the $\rho^{\pm}$ signal and
absence of the $\rho^{0}$ signal lead us to the natural assumption that
$\rho$ mesons originate from an isospin 1 resonance. 
For the case of the resonance with $I=0$ 
one expects production of both neutral and charged $\rho$'s. 

Possible candidates from the list of the processes 
(\ref{eq1})-(\ref{eq7}) are 
$a_{1}(1260)$,
$a_{2}(1320)$ and
$\pi(1300)$.
These resonances have different spin and parity resulting in different
angular distributions for a recoil pion (Table~\ref{reslist}).

\begin{table}[hbtp]
\caption{List of $I=1$ resonances with expected angular distributions}
\label{reslist}
\centering 
\renewcommand{\arraystretch}{1.1}
\begin{tabular}[t]{|c|l|c|l|r|}
\hline
& & $J^{PC}$ & 
$d\sigma/d\cos(\theta)$ &
$\chi^2/ndf$ \\
\hline
\hline
1 &
$a_{1}(1260)$  &
$1^{++}$  &
$const$ &
$7.1/7$\\
\hline
2 &
$\pi(1300)$ &
$0^{-+}$ &
$\sin^{2}(\theta)$ &
$47.3/7$\\
\hline
3 &
$a_{2}(1320)$ &
$2^{++}$ &
$1+\cos^{2}(\theta)$ &
$37.0/7$\\
\hline
\end{tabular}
\end{table}

\begin{figure}[htbp]
\centering 
\epsfig{figure=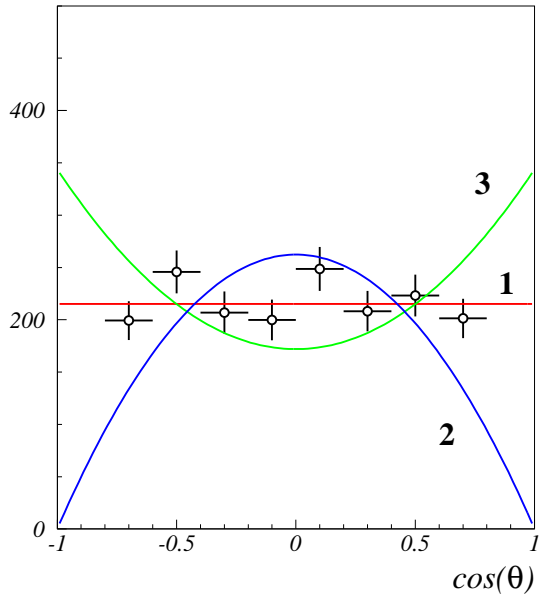,width=0.7\textwidth}
\caption{
Angular distribution of the recoil $\pi^{\pm}$.
Solid curves correspond to the following hypotheses: 
{\bf 1} -- $a_1(1260)\pi$;
{\bf 2} -- $\pi(1300)\pi$;
{\bf 3} -- $a_2(1320)\pi$
}
\label{picosfit}
\end{figure}

To measure the angular distribution of the $\rho^{\pm}\pi^0$ system 
(or recoiled $\pi^{\mp}$) we fit the $\rho^{\pm}$ signal in
eight ranges of the recoil $\pi^{\mp}$ angle with respect to the beam axis.
The measured angular distribution is shown in Fig.~\ref{picosfit}.
Three smooth curves show the fits corresponding to mentioned above
hypotheses. The fit goodness is presented in the last column
of Table~\ref{reslist}.
One can see good agreement with the hypothesis of the
$\anpi$ intermediate state.

Thus one can assume that our data can be explained by
$\ompi$ and $\anpi$ intermediate states although small
admixture of other states is not excluded.

\subsection{Fitting method}

For detailed comparison
we perform the simulation of the processes 
(\ref{eq1})-(\ref{eq7}) taking into account
their interference as well as the interference of the diagrams differing
by permutations of identical final pions (see Appendix \ref{appa}). 

In order to extract a relative fraction of 
$\anpi$ and any other intermediate
states, the unbinned maximum likelihood fit has been used \cite{kopylov}. 
Kinematics
of each event is completely described by 
a set of measured four-momenta 
$p_{\pi}\equiv (\varepsilon , \vec{\mathbf{p}})$
of the final pions with the exception of smearing due to detector resolution
and radiative effects. 
This set will be referred to as a vector of state 
$\vec{P}_{i}\equiv (p_{\pi^+},p_{\pi^-},p_{\pi^0},p_{\pi^0})_{i}$
where a subscript $i$ stands for the event number. 

The theoretical probability density function 
for $\vec{P}_{i}$ depends
on process dynamics and can be expressed via the matrix element $\mel$.
Here $\vec{A}$ stands for a set of unknown model parameters like relative
strength and phase of interference between the intermediate states. 
To obtain
the optimal values of $\vec{A}$ we minimize \cite{minuit} the logarithmic
likelihood function 
\begin{equation}
\label{lfunc}
L(\vec{A})=-\sum _{i=1}^{N_{2}}\ln \wfun \, ,
\end{equation}
 where $\wfun$ is the probability to observe an event in 
state $\vec{P}_{i}$,
and $N_{2}$ is the number of events in the second class. Since the detector
resolution for the invariant mass of three pions is about 10 MeV, 
i.e. comparable
to the $\omega$ width, smearing effects are significant for the events
of the first class. 
Therefore we use the events of the second class only. To
fix the fraction of events in the first class 
we add the following term to the function (\ref{lfunc})
\[
\frac{(r(\vec{A})-r_{1})^{2}}{2\sigma _{r_{1}}^{2}} \, ,
\]
 where $r_{1}=\frac{N_{1}}{N_{1}+N_{2}}$ is a measured fraction, 
$N_{1}$
is the number of events in the first class, 
and $r(\vec{A})$ is the expectation
of $r_{1}$.

The normalized probability density function 
is expressed via the matrix element
\[
\wfun =\frac{\mel}{\sigma_{vis}(\vec{A})}
\cdot\frac{1}{\prod_{j}2\varepsilon_{j}} \, ,
\]
where $j$ is the particle number, 
and $\sigma_{vis}(\vec{A})$ is
a model dependent visible cross section:
\[
\sigma _{vis}(\vec{A})=\int\mel
\cdot\prod_{j=1}^{4}\frac{d^{3}\vec{\mathbf{p}}_{j}}{2\varepsilon_j} \, .
\]

The function $\sigma_{vis}(\vec{A})$ is 
calculated using Monte Carlo technique.
For this purpose we use a set of events 
sampled according to the relativistic
phase space distribution. 
In this case, $\sigma_{vis}(\vec{A})$ is calculated
as
\[
\sigma_{vis}(\vec{A})=\frac{1}{N_{MC}}\sum_{i=1}^{N_{MC}}\mel \, , 
\]
where $N_{MC}$ is the number of second class events in a generated set. 

The main goal of our fits was to find the minimal model compatible with our
data. To this end, we write the matrix element as follows: 
\[
\mel =|\jel{\ompi }+Z_{a_{1}}\cdot \jel{\api }+Z_{X}\cdot \jel{X}|^{2} \, ,
\]
where $X$ stands for an admixture under study --- $\sigrho$, $\hnpi$,
$\rhorho$, $\aanpi$ or $\pinpi$. In the above equation,
expressions 
for $\jel{X}$ are taken from Appendix \ref{appa} while complex factors
$Z$ are components of the vector $\vec{A}$.

\subsection{\boldmath Comparison of $\pne$ data with simulation}

\begin{figure}[htbp]
\epsfig{figure=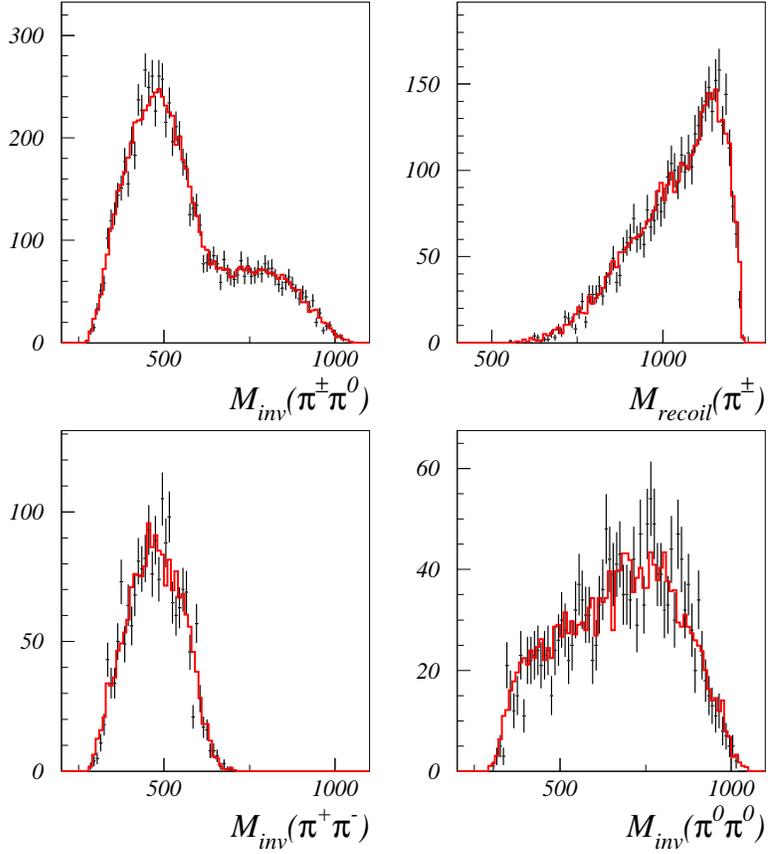,width=1.0\textwidth}
\caption{
Distributions over
$M_{inv}(\pi^{\pm}\pi^{0})$, $M_{recoil}(\pi^{\pm})$,
$M_{inv}(\pi^{+}\pi^{-})$, 
$M_{inv}(\pi^{0}\pi^{0})$
for $\pne$ events from the class (\ref{class1})
}
\label{omslide1}
\end{figure}

\begin{figure}[htbp]
\epsfig{figure=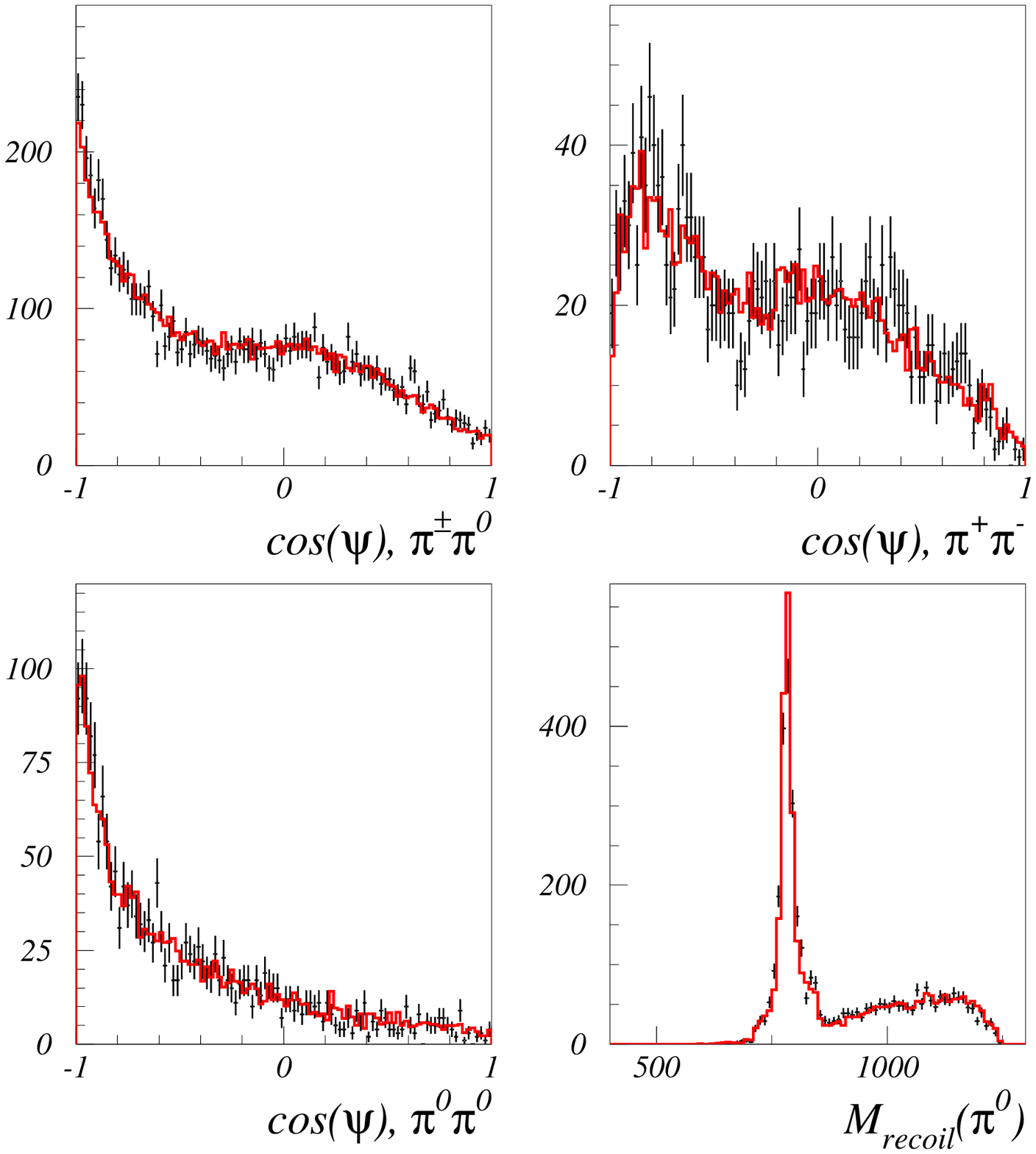,width=1.0\textwidth}
\caption{
Distributions over
$cos(\psi_{\pi^{\pm}\pi^{0}})$, $cos(\psi_{\pi^{+}\pi^{-}})$, 
$cos(\psi_{\pi^{0}\pi^{0}})$ and $M_{recoil}(\pi^{0})$
for $\pne$ events from the class (\ref{class1})
}
\label{omslide3}
\end{figure}

Since the matrix element for the channel $\ompi$  has a well-known
structure, one can use the events from the first class to test
the adequacy of the total MC simulation \cite{cmd2sim}.
Figs.~\ref{omslide1},\ref{omslide3} show the distributions over
$M_{inv}(\pi^{\pm}\pi^{0})$, $M_{recoil}(\pi^{\pm})$,
$M_{inv}(\pi^{+}\pi^{-})$, 
$M_{inv}(\pi^{0}\pi^{0})$, 
$cos(\psi_{\pi^{\pm}\pi^{0}})$, $cos(\psi_{\pi^{+}\pi^{-}})$, 
$cos(\psi_{\pi^{0}\pi^{0}})$ and $M_{recoil}(\pi^{0})$ for the events 
of the first class at the beam energy of 690 MeV.
Points with errors are the data, while the histograms 
correspond to the simulation of the 
processes $\ompi$ and $\anpi$.
Good consistence of the data and MC makes us confident that
MC simulation adequately reproduces both the kinematics of
produced particles and the detector response to them.

\begin{figure}[htbp]
\epsfig{figure=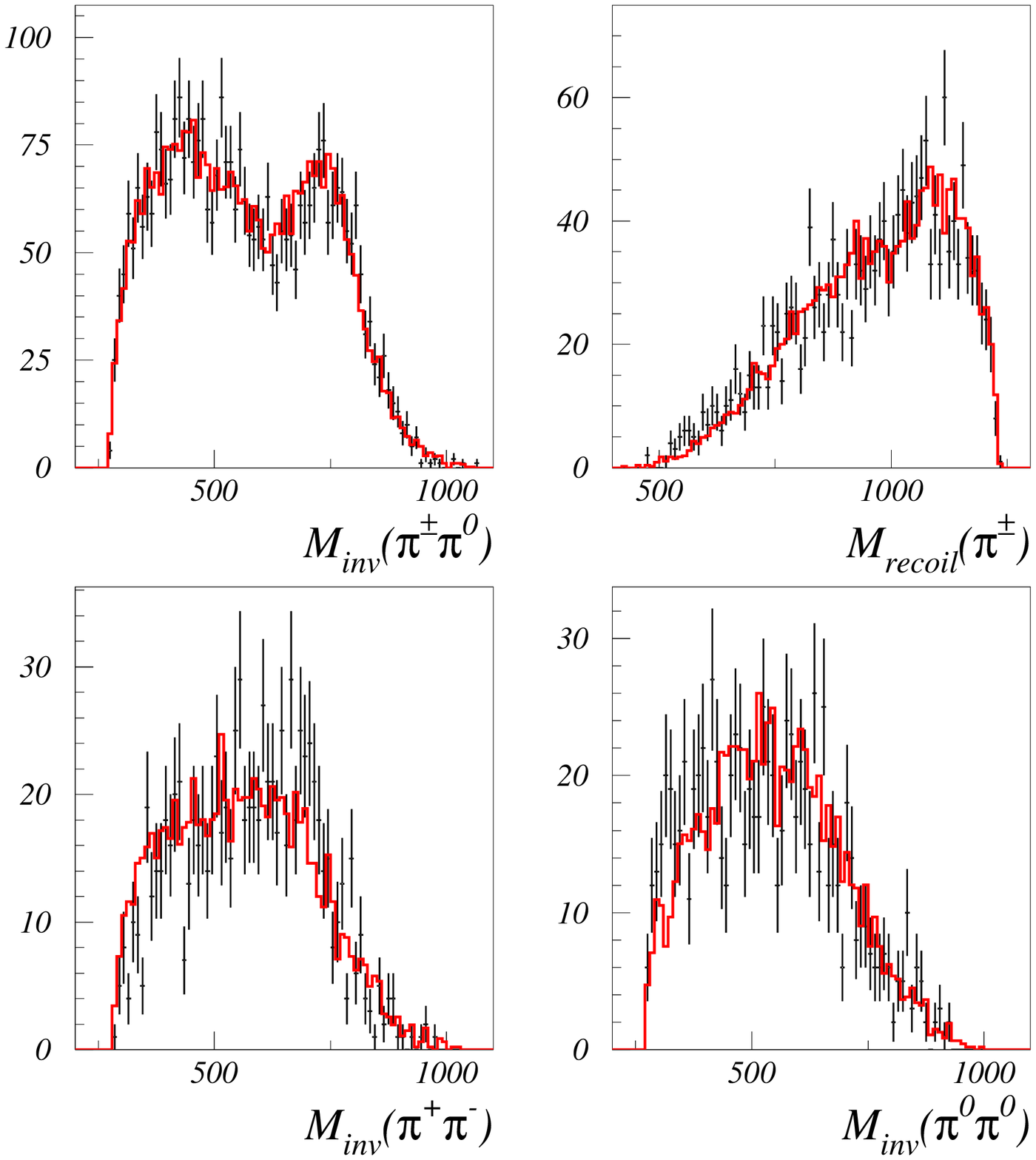,width=1.0\textwidth}
\caption{
Distributions over
$M_{inv}(\pi^{\pm}\pi^{0})$, $M_{recoil}(\pi^{\pm})$,
$M_{inv}(\pi^{+}\pi^{-})$, 
$M_{inv}(\pi^{0}\pi^{0})$
for $\pne$ events from the class (\ref{class2})
}
\label{a1slide1}
\end{figure}

\begin{figure}[htbp]
\epsfig{figure=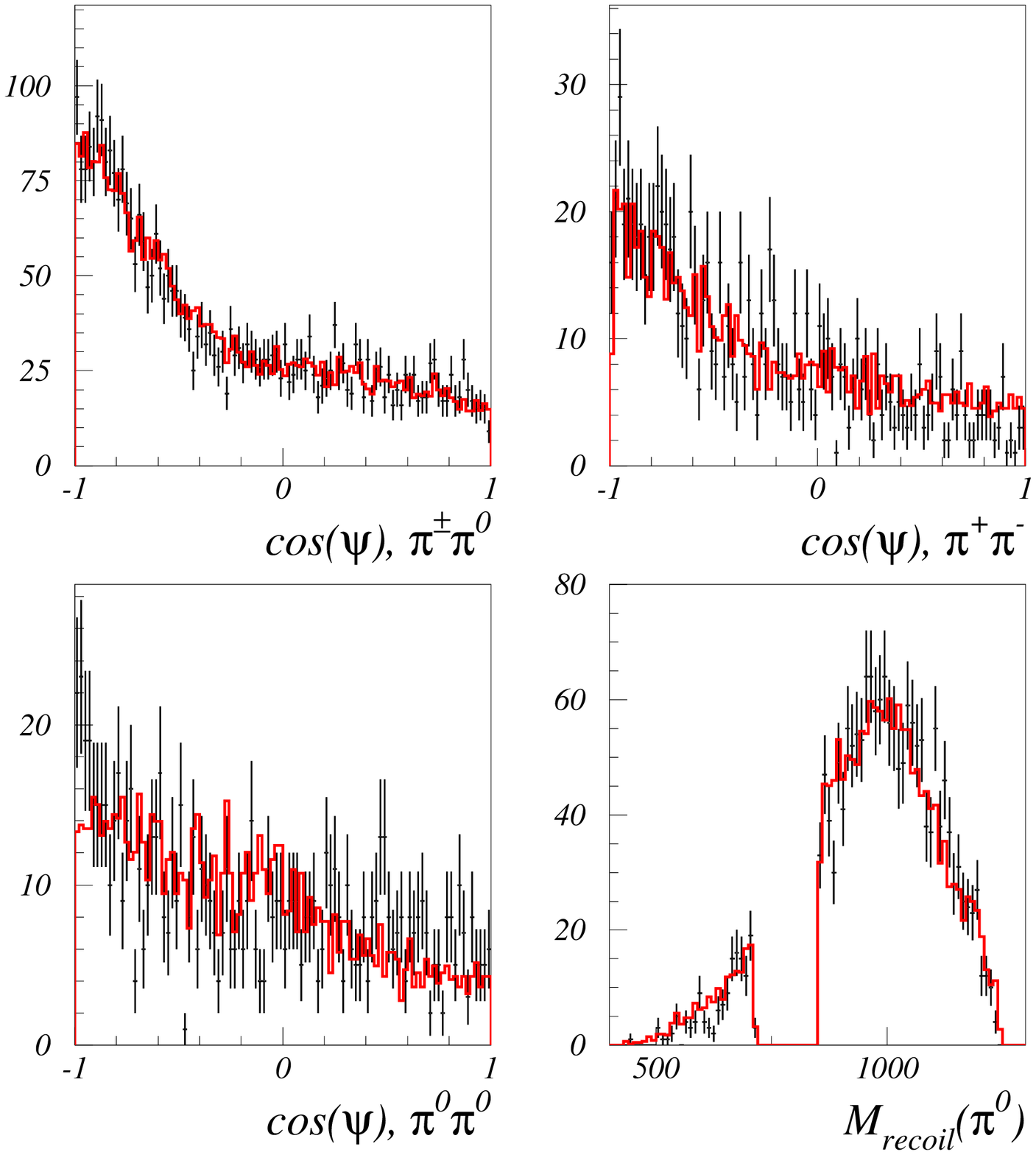,width=1.0\textwidth}
\caption{
Distributions over
$cos(\psi_{\pi^{\pm}\pi^{0}})$, $cos(\psi_{\pi^{+}\pi^{-}})$, 
$cos(\psi_{\pi^{0}\pi^{0}})$ and $M_{recoil}(\pi^{0})$
for $\pne$ events from the class (\ref{class2})
}
\label{a1slide3}
\end{figure}

Similar distributions for the events of the second class are shown in
Figs.~\ref{a1slide1},\ref{a1slide3}.
One can see that the process $\pne$ 
is satisfactorily described in the minimal model in which there are
two intermediate states $\ompi$ and $\anpi$ only.
Similar consistence is observed at other energies: we have also
examined
the energy points $2E_{beam}=(1.28\pm 0.01)$ and $(1.18\pm 0.03)$ GeV.

To determine the admixture of other possible mechanisms we extend the 
minimal model above by adding each of the other states one by one and 
performing the fit.
The results of these fits are shown in Table~\ref{admix}.
From its third column one can see that the relative fractions
$r_X$ of the additional intermediate state $X$ with respect to
the $\anpi$ are small. This
 confirms our assumption
of the $\anpi$ dominance.

\begin{table}[htbp]
\caption{The results of fit to different models with upper limits}
\label{admix}
\centering 
\renewcommand{\arraystretch}{1.4}
\begin{tabular}{|c|c|c|c|}
\hline 
Model&
\( L_{min}/N_{ev} \)&
$r_X$, $\%$ &
Upper limit, \(\%\)\\
\hline 
\hline 
\( \ompi +\api  \)&
\( 1264/891 \)&
---&
--- \\
\hline 
\( \ompi +\api +\sigrho  \)&
\( 1256/891 \)&
\( 2.1_{-0.9}^{+1.2} \)&
\( 4.3 \)\\
\hline 
\( \ompi +\api +\hpi  \)&
\( 1263/891 \)&
\( 0.1_{-0.1}^{+0.2} \)&
\( 0.4 \)\\
\hline 
\( \ompi +\api +\aapi  \)&
\( 1263/891 \)&
\( 0.2_{-0.2}^{+0.4} \)&
\( 0.8 \)\\
\hline 
\( \ompi +\api +\piprimpi  \)&
\( 1250/891 \)&
\( 9.5_{-2.8}^{+3.2} \)&
\( 15. \)\\
\hline 
\( \ompi +\api +\rhorho  \)&
\( 1246/891 \)&
\( 4.7_{-1.6}^{+2.0} \)&
\( 7.7 \)\\
\hline 
\end{tabular}
\end{table}

Because of the theoretical uncertainty of the
$\anpi$ matrix element, 
we do not consider the nonnegligible magnitude
of the $\rhorho$ and $\pinpi$ contributions as significant.
Instead, we prefer to set upper limits for the fraction
of these intermediate states.
The values of the likelihood function~(\ref{lfunc}) for
optimal parameters ($L_{min}$) do not contradict 
to our expectation based on the simulation of $\ompi$ and $\anpi$. 

\subsection{\boldmath Comparison of $\pch$ data with simulation}

Consider now the $\pch$ channel.
In this case we don't have $\ompi$ in the intermediate state
and therefore no additional complicacy to check the assumption
of the  $a_{1}(1260)\pi$ dominance arises.
To this end four-track events were selected.
All tracks should come from the interaction region: a distance from
track trajectories to the beam axis should be less than 1~cm and
the vertex position along the beam axis should be inside $\pm$ 15~cm.
After that a kinematic fit was performed assuming the $\pch$
hypothesis and events with $\chi^{2}/ndf~<~2.5$ were selected.
Under these conditions 28552 events remain 
in the energy range under study. 

\begin{figure}[htbp]
\epsfig{figure=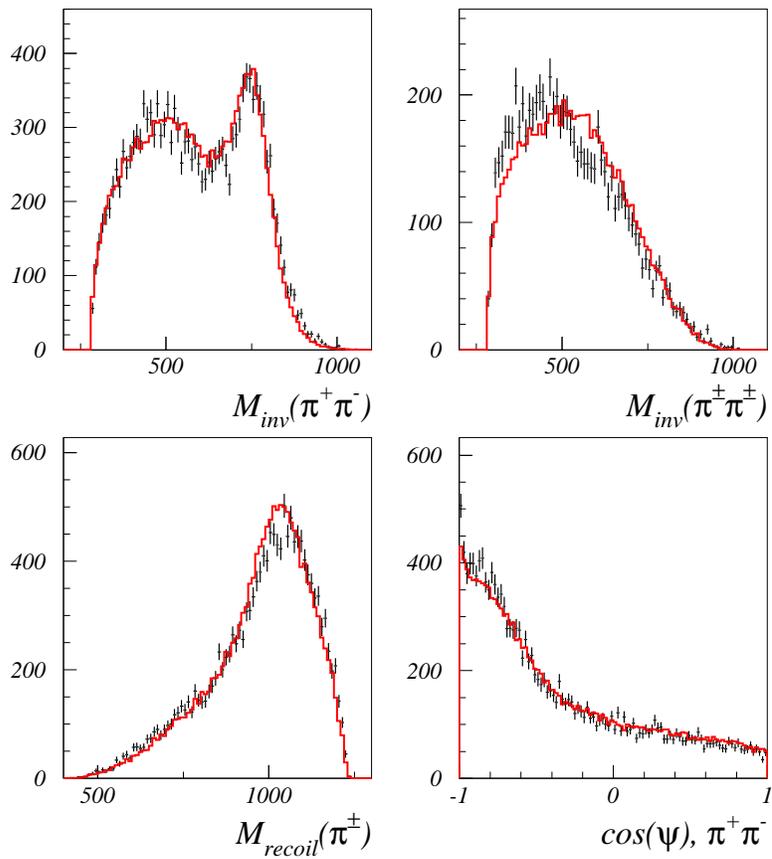,width=1.0\textwidth}
\caption{
Distributions over
$M_{inv}(\pi^{+}\pi^{-})$,
$M_{inv}(\pi^{\pm}\pi^{\pm})$,
$M_{recoil}(\pi^{\pm})$ and
$cos(\psi_{\pi^{+}\pi^{-}})$
for $4\pi^{\pm}$ events
}
\label{pi4slide}
\end{figure}

Figure~\ref{pi4slide} shows distributions over 
$M_{inv}(\pi^{+}\pi^{-})$,
$M_{inv}(\pi^{\pm}\pi^{\pm})$,
$M_{recoil}(\pi^{\pm})$ and
$cos(\psi_{\pi^{+}\pi^{-}})$ for \pch case.
One can see that the hypothesis of the $a_{1}(1260)\pi$ dominance does not
contradict to the data although one should note a slight deviation in the
spectrum of the invariant masses of the likesign pions.
This deviation can be possibly explained by the  contribution of the
$D-wave$ or some final state interaction. It can't be explained by 
the admixture of other possible processes  
because
their fractions compared to $a_{1}(1260)\pi$ are 
small (see Table~\ref{admix}). 
The study of 
more complicated cases taking into account simultaneously
a few intermediate states in addition to $\ompi$ and $\anpi$ 
is now in progress. 

\section{Energy dependence of the cross sections}

To obtain the $\anpi$ contribution to
the total cross section, we subtracted from the latter the contribution
of the $\ompi$ intermediate state. Strictly speaking, such a procedure
loses its meaning when the interference between $\ompi$ and $\anpi$
is strong. In our case, however, the effects of the interference
is numerically small ($\sim 5\%$) because of the small width of
the $\omega$ meson.

In order to diminish the influence of background from soft photons,
the additional
cut 
was applied:
$cos(\psi)<0.7$ 
 where $\psi$ is the angle between
the photon direction in the $\pi^{0}$ rest frame and $\pi^0$ momentum. 

Cross sections were calculated from the formulae:
\begin{eqnarray*}
\sigma_{\omega\pi^0} & = & \frac{N_{\omega}}
{B(\omega \to 3\pi)\cdot (1+\delta)\cdot L\cdot \epsilon} \; , \\
\sigma_{\pne} & = & \frac{N_{a_{1}}}
{(1+\delta)\cdot L\cdot \epsilon} \; , \\
\end{eqnarray*}
where 
 $L$ is the luminosity,
 $\epsilon$ is the detection efficiency from MC, and
 $\delta$ is a radiative correction calculated according to
\cite{kurfad}.


$N_{\omega}$ and $N_{a_{1}}$ are the numbers of events due to
the corresponding reactions ($\epm \to \ompi$ and $\epm \to \api$
respectively) under the assumption that there is no interference. 
The values of $N_{\omega}$ and $N_{a_{1}}$ were determined 
from the equations:
\begin{eqnarray*}
N_1 & = & N_{\omega}\cdot \alpha + N_{a_1}\cdot(1-\beta) \, , \\
N_2 & = & N_{\omega}\cdot(1-\alpha) + N_{a_1}\cdot\beta \, , \\
\end{eqnarray*}
where
$N_{1}$ and $N_{2}$ are the numbers of events 
in the first and second classes respectively,
$\alpha$ is the probability for the $\ompi$ event to enter 
the first class, and $\beta$ is the probability for the $\anpi$ 
event to enter the second class. The values of $\alpha$ and $\beta$ as 
a function of energy were taken from MC simulation.

\begin{figure}[htbp]
\centering
\epsfig{figure=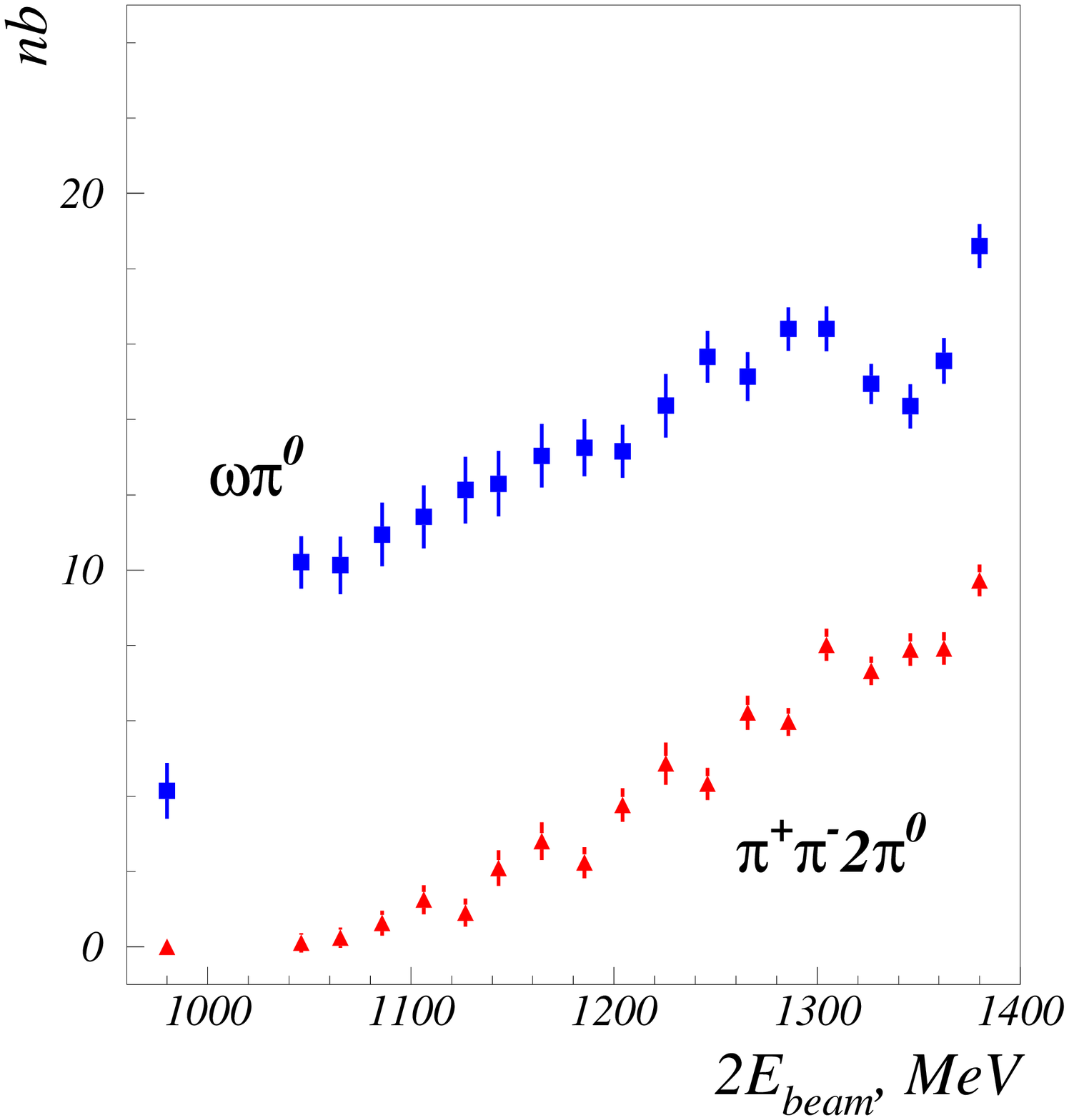,width=0.6\textwidth}
\caption{
Energy dependence of $\sigma(\epm \to \ompi)$ 
and $\sigma(\epm \to \pne)$ where
the
contribution of $\ompi$ is subtracted
}
\label{xsection}
\end{figure}

Figure~\ref{xsection} shows the cross sections obtained for
$\epm \to \ompi$ and
$\epm \to \pne$ with the $\ompi$ contribution subtracted as explained
above. Both cross sections rise with energy while the relative fraction
of the $\anpi$ increases.

\begin{figure}[htbp]
\centering
\epsfig{figure=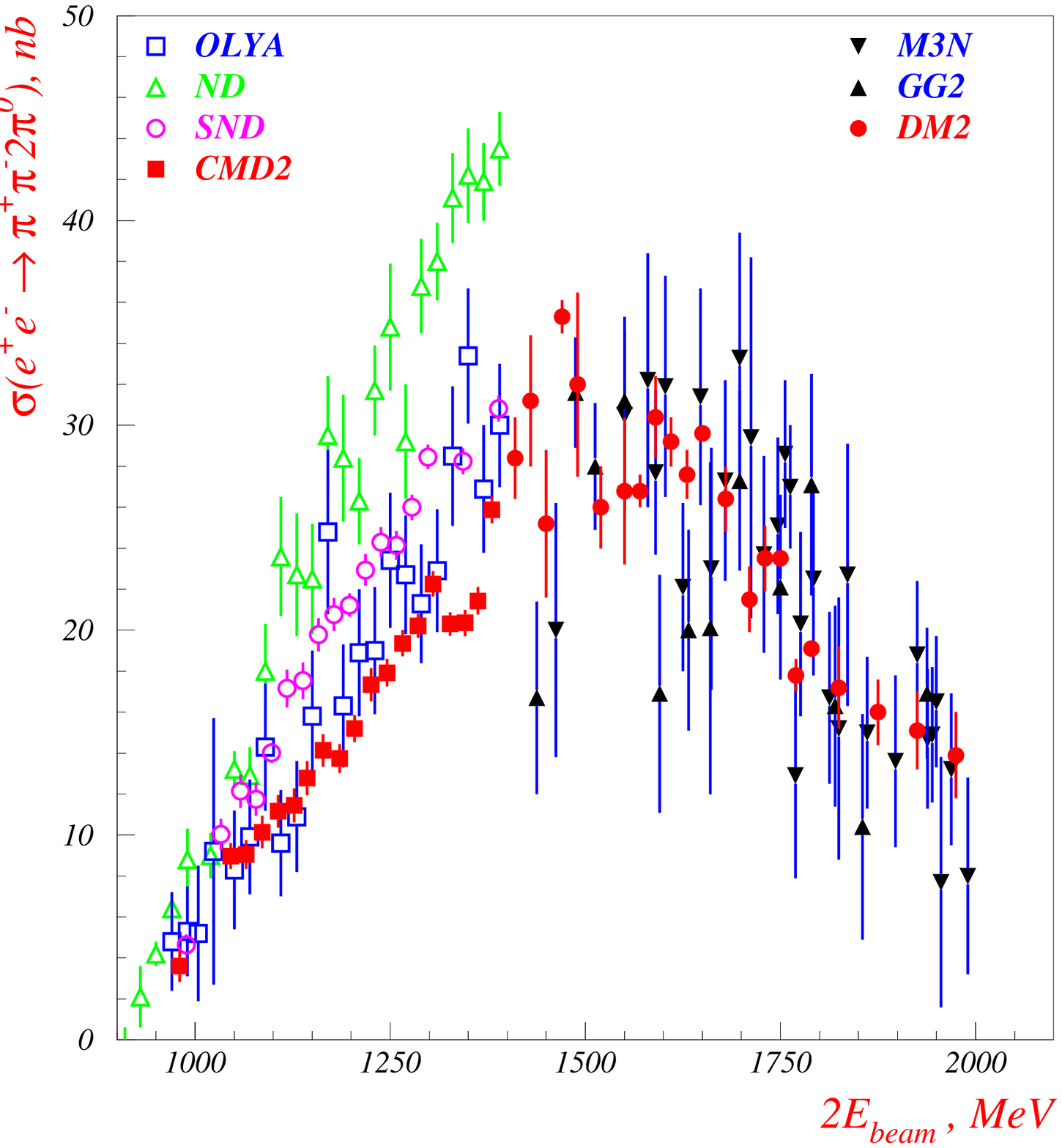,width=\textwidth}
\caption{
Energy dependence of the $\pne$
cross section}
\label{xs2pi}
\end{figure}

Figure~\ref{xs2pi} shows the total cross section vs energy. Only
statistical errors are shown. 
The systematic uncertainty for this channel consists of the contributions
from the uncertainties in the event reconstruction, radiative corrections
and luminosity determination. 
The overall systematic uncertainty
was estimated to be $\sim~15\%$.

The cross section measured in our experiment is consistent with the 
previous measurement at OLYA \cite{olya1} and within systematic errors does 
not contradict to the recent result from SND \cite{snd98}. The
value of the cross section from all three groups is significantly lower
than that from ND \cite{nd,ndr}. Also shown above 1.4 GeV are
results from Orsay \cite{dm2,m3n} and Frascati \cite{gg2} groups. 

To determine the cross section $\epm \to \pch$
the following criteria were added:
\begin{itemize}
\item $\chi^{2}/ndf~<~7$
\item $\theta_{min}~>~0.67$
\item $|\sum_{i} E_{i}~-~2\cdot E_{beam}|~<~0.2 \cdot E_{beam}$
\item $|\sum_{i} \overrightarrow{p_{i}}|~<~0.3 \cdot E_{beam}$
\end{itemize}
where $E_{i}$, $\overrightarrow{p_{i}}$ are pion energies and momenta 
before the kinematic reconstruction, and $\theta_{min}$ is a minimal
angle of pions with respect to the beam axis.

\begin{figure}[htbp]
\centering
\epsfig{figure=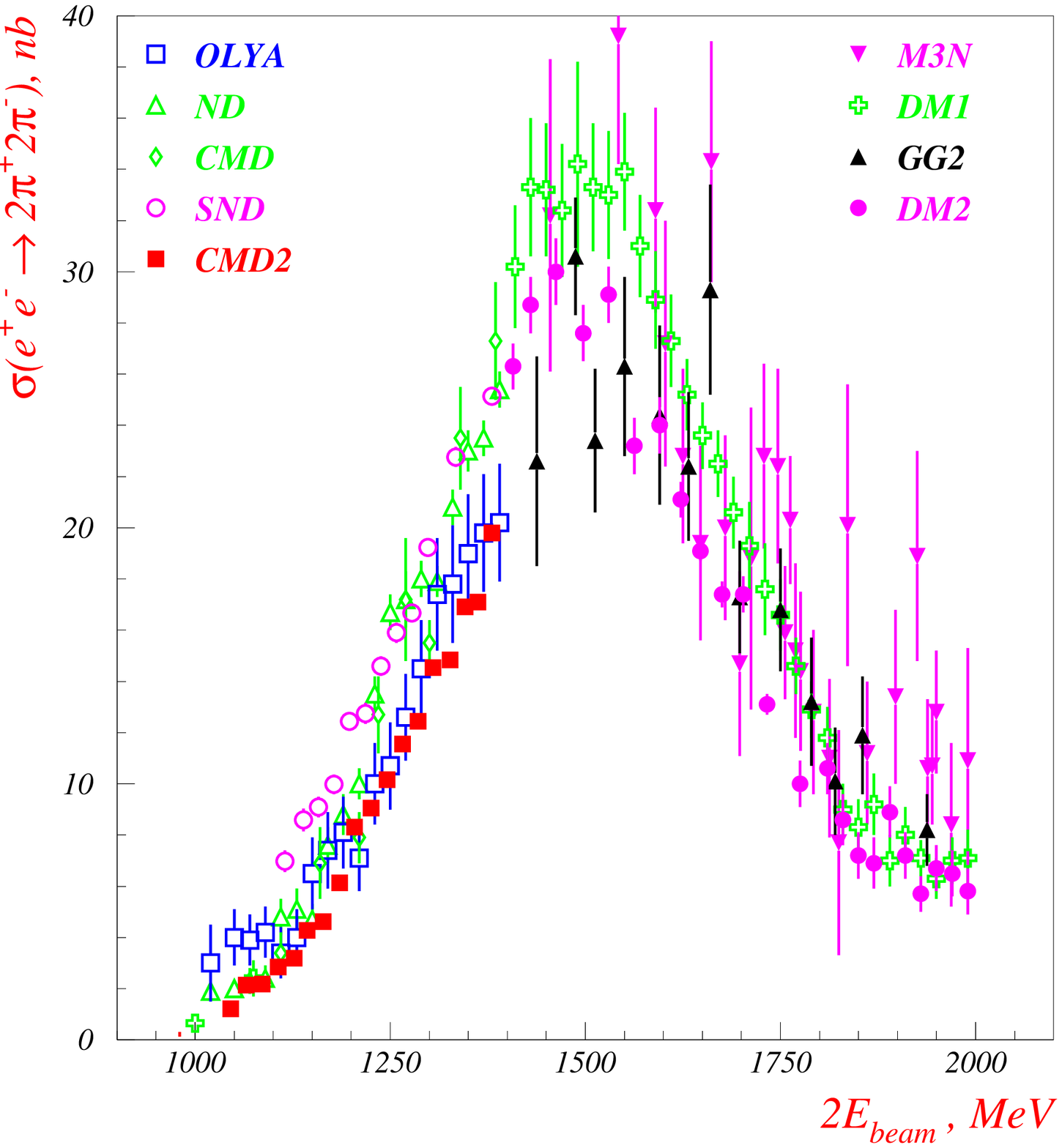,width=\textwidth}
\caption{
Energy dependence of the $\pch$
cross section}
\label{xs4pi}
\end{figure}

Figure~\ref{xs4pi} shows the total cross section vs energy. Only
statistical errors are shown. 
The systematic uncertainty for this channel consists of the contributions
from the uncertainties in the event reconstruction, radiative corrections,
selection criteria and luminosity determination. 
The overall systematic uncertainty
was estimated to be $\sim~15\%$.

The obtained cross section is 
consistent with the previous measurements at OLYA \cite{olya1},
ND \cite{nd,ndr} and CMD \cite{cmd} 
and within systematic errors does not contradict to the recent result 
from SND \cite{snd98}. Also shown above 1.4 GeV are results from
Orsay \cite{m3n,dm1,dm2} and Frascati \cite{mea,gg2:xs} groups.

\begin{figure}[htbp]
\centering
\epsfig{figure=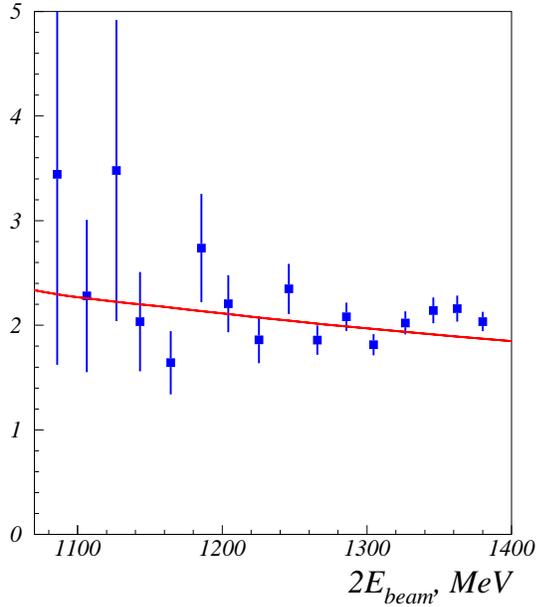,width=0.7\textwidth}
\caption{
Energy dependence of the ratio
$\frac{\sigma(\epm\to\pch)}
 {\sigma(\epm\to\pne)}$.
The
contribution of $\omega\pi^{0}$ is subtracted.
The solid curve shows the theoretical prediction 
based on the $a_{1}(1260)\pi$ dominance
}
\label{xsratio}
\end{figure}

Figure~\ref{xsratio} presents the ratio of the cross sections 
$\sigma(\epm \to \pch)$ and
$\sigma(\epm \to \pne)$ 
where the
contribution of $\omega\pi^{0}$ is subtracted.
The solid curve shows the theoretical prediction 
based on the $a_{1}(1260)\pi$ dominance.

In the energy range
under consideration the experimentally measured ratio is 
about two. Calculations show that this ratio tends to four at 
very low energies close to the threshold of four pion
production when the interference effects are maximal.
The effects of interference decreases with energy since 
the produced 
resonances are moving at a high velocity.
As a result, the ratio falls with energy and tends to unity.

The importance of the interference between the different diagrams
corresponding to permutations of identical
pions was first noted in \cite{ei}.
The neglection of interference can lead to wrong conclusions about
the contributions of different intermediate states to the total
cross section.

\section{Conclusion}

The detailed analysis of the process $\epm\to\pne$
unambiguously demonstrates that the dominant
contribution to the cross section comes from
$\ompi$ and $\rho^{\pm}\pi^{\mp}\pi^0$ intermediate states
whereas the  $\rho^0\pi^0\pi^0$ state is not observed.
Moreover, the $\rho^{\pm}\pi^{\mp}\pi^0$ state is completely
saturated by the $\anpi$ mechanism.
The latter also dominates in the cross section
of the process $\epm\to\pch$. This conclusion is based
on the data sample corresponding to
the integrated luminosity of 5.8 $pb^{-1}$
collected with the CMD-2 detector in the energy range
1.05 -- 1.38 GeV.

Comparison of the experimental data with the calculations
shows that
account of the interference of different amplitudes
with various intermediate resonances but
the identical final state
drastically changes predictions for the 
differential distributions and total
cross
sections.

Through the hypothesis of CVC \cite{CVC1} experimentally tested
to be valid within the 3-5$\%$ accuracy \cite{CVC2}, we can relate
the values of the four pion cross sections measured in $\epm$ to the     
hadronic spectra in corresponding four pion decays of the
$\tau$-lepton. Moreover, our observation of the $a_{1}(1260) \pi$
dominance should be also applied to $\tau$-lepton decays.

One should note that such a conclusion is in contradiction to that
made by ARGUS collaboration \cite{argus}. 
However, their conclusion based on
the distribution over the dipion mass is not compatible with
the recent data from ALEPH \cite{aleph}. The model used by ARGUS
didn't take into account the interference of the amplitudes
corresponding to various intermediate states whereas our
observation unambiguously points to significance of the
interference effects.  

Our predictions for various two- and three-body 
invariant mass spectra will be addressed in more detail 
in the forthcoming 
paper which will  compare them to the existing $\tau$-lepton data  
from ARGUS~\cite{argus}, CLEO~\cite{cleo} and ALEPH~\cite{aleph}.

The detailed investigation of the energy behavior
of exclusive channels of $\epm$ annihilation into
hadrons is crucial for understanding the structure of the
initial hadronic state created by the virtual photon
($\rho(1450)$, $\rho(1700)$ etc.).
Our work considers this problem in the energy range
below 1.4 GeV. 
At higher energy, such analysis can't be performed because
$\epm$ annihilation data are not available now.
One can hope to move forward by using
data on semihadronic decays of the $\tau$-lepton.

Our statement about the $\anpi$ dominance can be used
for verification of theoretical models in which
the problem of intermediate states has been discussed.
In particular, recently there were numerous speculations
about the possible existence of vector hybrids
claiming that hybrids can be distinguished
from the normal $q\bar{q}$ states by their decays, 
e.g. in the $\anpi$ \cite{hybrid}.

\section*{Acknowledgements}
The authors are grateful to the staff of VEPP-2M for excellent performance of 
the collider, to all engineers and technicians who participated in the design,
commissioning and operation of CMD-2.

Special thanks are due to N.N.~Achasov, M.~Benayoun,
V.L.~Chernyak, V.F.~Dmitriev, V.S.~Fadin, V.B.~Golubev,
I.B.~Khriplovich, L.~Montanet, S.I.~Serednyakov,
A.I.~Vainshtein for numerous discussions and 
constant interest.

\appendix
\renewcommand{\appendixname}{Appendix}
\begin{center}
\bf
\Large
Appendix
\end{center}

\section{\boldmath Models of substructure in $\epm\to 4\pi$}\label{appa}

In this section we discuss the details of the model which has
been used for description of the experimental data.
Since the initial hadron state (referred to as $\tilde\rho$) is created 
by a virtual photon  ($e^+e^- \to \gamma^*\to\tilde\rho $) and then decays
into  four pions, it has the $\rho$-meson quantum numbers
($I^{G}J^{PC}=1^{+}1^{--}$). We use the notation  ${\tilde e}_{\mu}$
for the polarization vector of this state. 

We assume that the main contribution to the amplitude of the process
in the energy range under study
is given by the intermediate state  resonances having the masses
close to the 
threshold of $\rho\pi$ production (\ref{eq1})-(\ref{eq7}).
In the case of broad resonances ($a_1(1260)$,$a_2(1320)$,$\pi(1300)$ etc.),
the form of their propagators $1/D(q)$ 
is very important for 
analysing the data.
We learned  that it is necessary to take into account
the dependence of the imaginary part of the propagators (width)
on virtuality while
the corresponding corrections to the real part which can be expressed
through the imaginary part by dispersion relations are not
so important. 

We represent the function $D(q)$ 
in the form
\beq
D(q)=q^2-M^2 +iM\Gamma \frac{g(q^2)}{g(M^2)} \, ,
\eeq
where $M$ and $\Gamma$  are the mass and width of the corresponding
particle, and the function $g(s)$ describes the dependence of the width
on the virtuality.
If $q^2=M^2$ then $D=iM\Gamma$,  in accordance with the usual
definition of  mass and  width of the resonance.
In the case of the $\rho$-meson we used the following representation for
the function $g_{\rho}(q)$: 
\beq
g_{\rho}(s)=s^{-1/2}(s-4m^2)^{3/2}  \, .
\eeq

\subsection{The contribution of $a_1(1260)\pi$}

The quantum numbers of the $a_1(1260)$ resonance are
$I^{G}J^{PC}=1^{-}1^{++}$.
Taking into account the quantum numbers of the pion, we can write 
the matrix elements corresponding to the processes
$\tilde\rho(P)\to a_1(q)\pi(p)$ and
$a_1(q)\to\rho(P^{\prime})\pi(p)$ as
\beqn\label{M1}
T(\tilde\rho\to a_1\pi)=F_{\tilde\rho a_1\pi}\eps^{3ab}
(P_{\mu}{\tilde e}_{\nu}-P_{\nu}{\tilde e}_{\mu})
q_{\mu}A_{\nu}^{a*}\phi^{b*}\, ,\nonumber\\
T(a_1\to \rho\pi)=F_{a_1\rho\pi}\eps^{abc}
q_{\mu}A_{\nu}^{a}
(P_{\mu}^{\prime}e_{\nu}^{b*}-P_{\nu}^{\prime}e_{\mu}^{b*})\phi^{c*} \, ,
\eeqn
where $a,b,c$ are isospin indices, $A_{\mu}^a$ and  $e_{\mu}^a$ are
the polarization vectors of $a_1$- and $\rho$-mesons, $\phi^b$ is the
pion wave function,
$F_{a_1\rho\pi}$ and $F_{\tilde\rho a_1\pi}$ are form factors
depending on the virtuality of initial and final particles. 
The explicit form of these form factors in the energy range considered
is not very essential (it contributes to the theoretical uncertainty
of the model). The matrix element of the transition 
 $T(\rho\to \pi\pi)$ reads
\beq\label{Mrho}
T(\rho\to \pi\pi)=F_{\rho\pi\pi}\eps^{abc}
e_{\mu}^{a}(p_{\mu}^{(b)}-p_{\mu}^{(c)})\phi^{b*}\phi^{c*}\, ,
\eeq
where $p_{\mu}^{(b)}$ and $p_{\mu}^{(c)}$ are 4-momenta of the corresponding
pions. Due to the helicity conservation 
only transverse space components (with respect
to electron and positron momenta) of the 4-vector ${\tilde e}_{\mu}$ are 
not zero.
We denote them as $\tilde{\bf e}_{\perp}$. For unpolarized
electrons and positrons a square of matrix element
for the process $\tilde\rho \to 4\pi$ is of the form
\beq
\mid T\mid^2 =\, \mid {\bf J}_{\perp}\mid^2 .
\eeq
Using  \eq{M1} and  \eq{Mrho} we obtain the following expression for 
the contribution of the $a_1(1260)$-meson to the current  $ {\bf J}$ in
the process \\
$\tilde\rho\to \pi^+{(p_1)}\pi^+{(p_2)}\pi^-{(p_3)}\pi^-{(p_4)}$ :
\beqn\label{JAPPMM}
{\bf J}_{a_1}^{++--}\!\!\!&=&\!\!\!
G\left[\T_{a_1}(p_1,p_2,p_3,p_4)+\T_{a_1}(p_1,p_4,p_3,p_2)
+\T_{a_1}(p_2,p_1,p_3,p_4)
\right. \nonumber\\
&&+\T_{a_1}(p_2,p_4,p_3,p_1)+\T_{a_1}(p_1,p_2,p_4,p_3)
+\T_{a_1}(p_1,p_3,p_4,p_2) \nonumber \\
&&+\left.\T_{a_1}(p_2,p_1,p_4,p_3)
+\T_{a_1}(p_2,p_3,p_4,p_1)\right] \, ,  
\eeqn
where
\beqn\label{j} 
\dst
\T_{a_1}(p_1,p_2,p_3,p_4)=\frac{F_{a_1}^2(P-p_4)}
{D_{a_1}(P-p_4)D_{\rho}(p_1+p_3)}\\
\dst
\times\{(E-\eps_4)[\p_1(E\eps_3-p_4p_3)-\p_3(E\eps_1-p_4p_1)] \nonumber \\
\dst
- \p_4[\eps_1(p_4p_3)-\eps_3(p_4p_1)]\}\, ,
\nonumber
\eeqn
$P$ is the initial 4-momentum
( $P^0=E$ , ${\bf P}=0$)  , $p_i=(\eps_i, \p_i)$ ,
 $1/D_A(q)$ and   $1/D_{\rho}(q)$ are propagators of the 
$a_1$-  and $\rho$- mesons, $F_{a_1}(q)$ is the 
form factor, $G$ is the function
of the initial energy $E$ proportional to the amplitude of 
$\tilde\rho$ creation.
Similarly to \eq{JAPPMM}, we obtain for the contribution of
$a_1(1260)$ to the current $ {\bf J}$ in the process
$\tilde\rho\to \pi^+{(p_1)}\pi^-{(p_4)}\pi^0{(p_2)}\pi^0{(p_3)}$:
\beqn\label{JAPM00}
{\bf J}_{a_1}^{+-00}&=&G\left[\T_{a_1}(p_1,p_2,p_3,p_4)
-\T_{a_1}(p_4,p_2,p_3,p_1)\right. \\
&&+\left.\T_{a_1}(p_1,p_3,p_2,p_4)-\T_{a_1}(p_4,p_3,p_2,p_1)\right]\, .
\nonumber  
\eeqn 

The function $g_{a_1}(s)$ in the propagator of $a_1$ reads: 
\beqn\label{g} 
g_{a_1}(s)&=& F_{a_1}^2(Q) \int \left | 
 \frac{\eps_2\p_1-\eps_1\p_2}{D_{\rho}(p_1+p_2)}+
 \frac{\eps_2\p_3-\eps_3\p_2}{D_{\rho}(p_2+p_3)}\right|^2 \nonumber \\
&& \times \frac{d\p_1\,d\p_2\,d\p_3\,\delta^{(4)}(p_1+p_2+p_3-Q)}
 {2\eps_12\eps_2 2\eps_3(2\pi)^5}\, ,
\eeqn
where $Q^0=\sqrt s$ and ${\bf Q}=0$.  As a form factor, we used
the function $F(q)=(1+m_{a_1}^2/\Lambda^2)/(1+q^2/\Lambda^2)$ 
with $\Lambda\sim$ 1~GeV. 

\subsection{The contribution of $\omega\pi$}

The amplitude of the process $\tilde\rho({\cal P})\to \omega(q)\pi(p)$
has the form:
\beq\label{om}
T(\tilde\rho\to \omega\pi)=F_{\tilde\rho\omega\pi}
\eps_{\mu\nu\alpha\beta}{\cal P}_{\mu}
q_{\nu}\tilde{e}_{\alpha}^a e_{\beta}^a\, ,
\eeq
where $ e_{\beta}^a$ is the polarization vector of the $\omega$-meson.
The matrix element of the transition $\omega\to\rho\pi$
can be written in the similar form. The  $\omega$-meson
contributes only to a channel  
$\tilde\rho\to \pi^+{(p_1)}\pi^-{(p_4)}\pi^0{(p_2)}\pi^0{(p_3)}$ .
The corresponding current is equal to
\beqn\label{JOMPM00}
{\bf J}_{\omega}^{+-00}&=&
G_{\omega}\left[\T_{\omega}(p_2,p_4,p_1,p_3)-
\T_{\omega}(p_2,p_1,p_4,p_3)\right.\nonumber\\
&&-\left.\T_{\omega}(p_2,p_3,p_1,p_4)
\right] + (p_2\leftrightarrow p_3) \, ,  
\eeqn 
where
\beqn 
\dst
\T_{\omega}(p_1,p_2,p_3,p_4)=
\frac{F_{\omega}^2(P-p_1)}{D_{\omega}(P-p_1)D_{\rho}(p_3+p_4)}\\
\dst
\times \{(\eps_4\p_3-\eps_3\p_4)(\p_1\p_2)-
\p_2(\eps_4\p_1\p_3-\eps_3\p_1\p_4) \nonumber \\
\dst
-\eps_2[\p_3(\p_1\p_4)-\p_4(\p_1\p_3)]\}
\, , \nonumber
\eeqn
$\eps_i$ is the energy of the corresponding pion, $F_{\omega}(q)$ is
the form factor.
Since 
the width of the $\omega$ is small,
we set $g_{\omega}(s)=1$
in the propagator $D_{\omega}(q)$. 

\subsection{The contribution of $h_1(1170)\pi$} 

The quantum numbers of the $h_1(1170)$ resonance are $I^{G}J^{PC}=0^{-}1^{+-}$.
Similarly to the $\omega$-meson it gives the contribution to the mixed 
channel only.
The matrix elements for the transitions $\tilde\rho(P)\to h_1(q)\pi^0(p)$ and
 $h_1(q)\to\rho(P)\pi(p)$ have the form
\beqn\label{H1}
T(\tilde\rho\to h_1\pi^0)=F_{\tilde\rho h_1\pi}
(P_{\mu}{\tilde e}_{\nu}-P_{\nu}{\tilde e}_{\mu})
q_{\mu}H_{\nu}^*\phi^{3*}\, ,\nonumber\\
T(h_1\to \rho\pi)=F_{h_1\rho\pi}
q_{\mu}H_{\nu}
(P_{\mu}e_{\nu}^{a*}-P_{\nu}e_{\mu}^{a*})\phi^{a*} \, ,
\eeqn
where  $H_{\mu}$ is the polarization vector of $h_1$.
Using \eq{H1} and \eq{Mrho} we obtain the following representation for 
the $h_1$-meson contribution to the current $ {\bf J}$ in the
decay $\tilde\rho\to \pi^+{(p_1)}\pi^-{(p_4)}\pi^0{(p_2)}\pi^0{(p_3)}$ :

\beqn\label{JHPM00}
{\bf J}_{h_1}^{+-00}&=&
G_{h_1}\left[\T_{h_1}(p_1,p_4,p_3,p_2)-
\T_{h_1}(p_4,p_1,p_3,p_2)\right.\nonumber\\
&&-\left.\T_{h_1}(p_1,p_3,p_4,p_2)\right] + (p_2\leftrightarrow p_3) \, ,
\eeqn
the current  $\T_{h_1}$ is given by  \eq{j} with the change of indices
$a_1\to h_1$ , and the function $g_{h_1}(s)$ in the propagator of $h_1$ is:
\beqn\label{gh} 
g_{h_1}(s)&=&F_{h_1}^2(Q) \int
\left | \frac{\eps_2\p_1-\eps_1\p_2}{D_{\rho}(p_1+p_2)}-
\frac{\eps_2\p_3-\eps_3\p_2}{D_{\rho}(p_2+p_3)}-
\frac{\eps_3\p_1-\eps_1\p_3}{D_{\rho}(p_1+p_3)}
 \right|^2 \nonumber\\
&& \times \frac{d\p_1\,d\p_2\,d\p_3\,\delta^{(4)}(p_1+p_2+p_3-Q)}
 {2\eps_12\eps_2 2\eps_3(2\pi)^5}\, .
\eeqn

\subsection {The contribution of $\rho^+\rho^-$}

One more contribution to the mixed channel amplitude comes from 
the process $\tilde\rho\to \rho^+\rho^-\to 4\pi$.
The matrix element corresponding to the transition 
$\tilde\rho(P)\to \rho^+(p)\rho^-(q)$ reads
\beqn\label{rhorho}
T(\tilde\rho\to \rho^+\rho^-)&=&F_{\tilde\rho \rho^+\rho^-}
(P_{\mu}{\tilde e}_{\nu}-P_{\nu}{\tilde e}_{\mu}) \\
&& \times [(p_{\mu}e_{\alpha}^{+*}-p_{\alpha}e_{\mu}^{+*})
(q_{\nu}e_{\alpha}^{-*}-q_{\alpha}e_{\nu})^{-*})-(\mu\leftrightarrow\nu)]
\nonumber
\eeqn
where  $e_{\mu}^+$ and  $e_{\mu}^-$ are the polarization vectors
of $\rho^+$ and $\rho^-$ respectively.
Using \eq{rhorho}, we obtain the contribution of   
 $\rho^+\rho^-$ to the current $ {\bf J}$ in the decay
$\tilde\rho\to \pi^+{(p_1)}\pi^-{(p_4)}\pi^0{(p_2)}\pi^0{(p_3)}$ :

\beqn\label{JRRPM00}
{\bf J}_{\rho\rho}^{+-00}&=&
\frac{G_{\rho\rho}F_{\rho}^2(p_1+p_2)F_{\rho}^2(p_3+p_4)}
{D_{\rho}(p_1+p_2)D_{\rho}(p_3+p_4)} \\
&& \times [\p_1(\eps_3\p_2\p_4-\eps_4\p_2\p_3)-
\p_2(\eps_3\p_1\p_4-\eps_4\p_1\p_3)\nonumber \\
&& -\p_3(\eps_1\p_2\p_4-\eps_2\p_1\p_4)+
\p_4(\eps_1\p_2\p_3-\eps_2\p_1\p_3)]   + (2\leftrightarrow 3)\, .
\nonumber
\eeqn

\subsection{The contribution of $\pi(1300)\pi$}

The matrix elements for the transitions 
$\tilde\rho(P)\to \pi^{\prime}(q)\pi(p)$ and
$\pi^{\prime}(q)\to\rho(P)\pi(p)$ are of the form
(for brevity $\pi^{\prime}\equiv\pi(1300)$):
\beqn\label{pp}
T(\tilde\rho\to \pi^{\prime}\pi)=F_{\tilde\rho \pi^{\prime}\pi}\eps^{3ab}
(P_{\mu}{\tilde e}_{\nu}-P_{\nu}{\tilde e}_{\mu})
q_{\mu}p_{\nu}\phi^{\prime a*}\phi^{b*}\, ,\nonumber\\
T(\pi^{\prime}\to \rho\pi)=F_{\pi^{\prime}\rho\pi}\eps^{abc}
q_{\mu}p_{\nu}
(P_{\mu}e_{\nu}^{b*}-P_{\nu}e_{\mu}^{b*})\phi^{\prime a}\phi^{c*} \, .
\eeqn

Then we get for the contribution of
$\pi(1300)$ to the current $ {\bf J}$ in the decay
$\tilde\rho\to \pi^+{(p_1)}\pi^+{(p_2)}\pi^-{(p_3)}\pi^-{(p_4)}$ :

\beqn\label{JPiPPMM}
{\bf J}_{\pi^{\prime}}^{++--}&=&
G_{\pi^{\prime}}\left[\T_{\pi^{\prime}}(p_1,p_2,p_3,p_4)+
\T_{\pi^{\prime}}(p_2,p_1,p_3,p_4)+\T_{\pi^{\prime}}(p_1,p_2,p_4,p_3)
\right. \nonumber\\
&&+\T_{\pi^{\prime}}(p_2,p_1,p_4,p_3)+\T_{\pi^{\prime}}(p_3,p_2,p_1,p_4)
+\T_{\pi^{\prime}}(p_4,p_2,p_1,p_3)\nonumber\\
&&+\left.\T_{\pi^{\prime}}(p_3,p_1,p_2,p_4)+\T_{\pi^{\prime}}
(p_4,p_1,p_2,p_3)\right]\, .
\eeqn
In the mixed channel 
$\tilde\rho\to\pi^+{(p_1)}\pi^-{(p_4)}\pi^0{(p_2)}\pi^0{(p_3)}$
the corresponding current reads:
\beqn\label{JPiPP00}
{\bf J}_{\pi^{\prime}}^{+-00}&=&
G_{\pi^{\prime}}\left[
\T_{\pi^{\prime}}(p_1,p_2,p_3,p_4)+\T_{\pi^{\prime}}(p_1,p_3,p_2,p_4)\right. 
\nonumber\\
&&+\left.\T_{\pi^{\prime}}(p_4,p_1,p_3,p_2)+\T_{\pi^{\prime}}(p_4,p_1,p_2,p_3)
\right] .
\eeqn
It \eq{JPiPPMM} and \eq{JPiPP00} we use the notation
\beq
\T_{\pi^{\prime}}(p_1,p_2,p_3,p_4)=
\frac{F_{\pi^{\prime}}^2(P-p_1)}{D_{\pi^{\prime}}(P-p_1)D_{\rho}(p_2+p_4)}
(p_2p_3-p_4p_3)(p_2+p_4)^2\,\p_1 \, .
\eeq
For the function $g_{\pi^{\prime}}(s)$ in the 
$\pi^{\prime}$ propagator, one has:
\beqn\label{gpi} 
g_{\pi^{\prime}}(s)&=&F_{\pi^{\prime}}^2(Q) \int
\left | \frac{\eps_1\p_2\p_3-\eps_2\p_1\p_3}{D_{\rho}(p_1+p_2)}+
\frac{\eps_3\p_1\p_2-\eps_2\p_1\p_3}{D_{\rho}(p_2+p_3)}
 \right|^2 \nonumber\\
&& \times \frac{d\p_1\,d\p_2\,d\p_3\,\delta^{(4)}(p_1+p_2+p_3-Q)}
 {2\eps_12\eps_2 2\eps_3(2\pi)^5}\, .
\eeqn

\subsection{The contribution of $\sigma\rho$}

The quantum numbers of the $\sigma$ resonance are $I^{G}J^{PC}=0^{+}0^{++}$.
The matrix element of the transition 
$\tilde\rho(P)\to \sigma(q)\rho^0(p)$ is of the form:
\beq\label{sigrho}
T(\tilde\rho\to \sigma\rho^0)=F_{\tilde\rho \sigma\rho}
(P_{\mu}{\tilde e}_{\nu}-P_{\nu}{\tilde e}_{\mu})
q_{\mu}e_{\nu}^*\phi_{\sigma}^{*} .
\eeq
The corresponding contribution to the current $ {\bf J}$ in the
decay \\
$\tilde\rho\to \pi^+{(p_1)}\pi^+{(p_2)}\pi^-{(p_3)}\pi^-{(p_4)}$ reads:
\beqn\label{SigRhoPPMM}
{\bf J}_{\sigma}^{++--}&=&
G_{\sigma}\left[\T_{\sigma}(p_1,p_2,p_3,p_4)+
\T_{\sigma}(p_2,p_1,p_3,p_4)\right. \nonumber\\
&&+\left.\T_{\sigma}(p_1,p_2,p_4,p_3)+\T_{\sigma}(p_2,p_1,p_4,p_3)\right]\, .
\eeqn
In the mixed channel
 $\tilde\rho\to\pi^+{(p_1)}\pi^-{(p_4)}\pi^0{(p_2)}\pi^0{(p_3)}$
the current has the form
\beq\label{SigRhoPP00}
{\bf J}_{\sigma}^{+-00}=G_{\sigma}\T_{\sigma}(p_1,p_2,p_3,p_4)\, ,
\eeq
where
\beq
\T_{\sigma}(p_1,p_2,p_3,p_4)=
\frac{F_{\sigma}^2(p_2+p_3)}{D_{\sigma}(p_2+p_3)D_{\rho}(p_1+p_4)}
(\eps_4\p_1-\eps_1\p_4) \, .
\eeq
The function $g_{\sigma}(s)$ in the propagator of $\sigma$ is
equal to
\beq
g_{\sigma}(s)=(1-4m^2/s)^{1/2}  \, .
\eeq

\subsection{The contribution of $a_2(1320)\pi$}

The quantum numbers of the $a_2(1320)$ 
resonance are $I^{G}J^{PC}=1^{-}2^{++}$.
The matrix elements for the transitions $\tilde\rho(P)\to a_2(q)\pi(p)$ and
 $a_2(q)\to\rho(P^{\prime})\pi(p)$ can be written in the form 
\beqn\label{a2}
T(\tilde\rho\to a_2\pi)=F_{\tilde\rho a_2\pi}\eps^{3ab}
\eps_{\mu,\nu\rho\lambda}P_{\mu}{\tilde e}_{\nu}p_{\rho}
A_{\lambda\gamma}^{a*}p_{\gamma}q_{\mu}\phi^{b*}\, ,\nonumber\\
T(a_2\to \rho\pi)=F_{a_2\rho\pi}\eps^{abc}
\eps_{\mu,\nu\rho\lambda}P_{\mu}^{\prime}{e}_{\nu}^{a*}p_{\rho}
A_{\lambda\gamma}^{b}p_{\gamma}\phi^{c*}\,
\eeqn
where $A_{\mu\nu}^a$ is the polarization tensor of $a_2$ .
The contributions of the  $a_2$-meson to the current $ {\bf J}$ in the
processes $\tilde\rho\to \pi^+{(p_1)}\pi^+{(p_2)}\pi^-{(p_3)}\pi^-{(p_4)}$
and $\tilde\rho\to \pi^+{(p_1)}\pi^-{(p_4)}\pi^0{(p_2)}\pi^0{(p_3)}$
are given by the formulae \eq{JAPPMM} and \eq{JAPM00} with
the substitution $\T_{a_1}\to\T_{a2}$ , where

\beqn\label{ja2} 
\dst
\T_{a2}(p_1,p_2,p_3,p_4)=\frac{F_{a2}^2(P-p_4)}
{D_A(P-p_4)D_{\rho}(p_1+p_3)}\\
\dst
\times\{E(\p_4\times\p_2)[\p_1\times\p_3,\p_2]+
[p_2p_4-(p_2q)(p_4q)/m_{a2}^2]  \nonumber\\
\dst
\times [\p_1(\eps_3\p_2\p_4-\eps_2\p_4\p_3)
-\p_3(\eps_1\p_4\p_2)-\eps_2\p_4\p_1) \nonumber \\
\dst
+\p_2(\eps_1\p_3\p_4-\eps_3\p_1\p_4) ]\}\, .
\nonumber
\eeqn
Here $q=P-p_4$, and $P$ is the initial 4-momentum.

For the function $g_{a2}(s)$ in the propagator of $a_2$, we obtain: 
\beq\label{ga2} 
g_{a2}(s)=F_{a2}^2(Q)\, \int {\cal F}_{ij} {\cal F}_{ij}^*
\frac{d\p_1\,d\p_2\,d\p_3\,\delta^{(4)}(p_1+p_2+p_3-Q)}
 {2\eps_12\eps_2 2\eps_3(2\pi)^5}\, ,
\eeq
where
\beq\label{gga2} 
 {\cal F}_{ij}(s)=\frac{(\p_2\times\p_3)^i\p_1^j}{D_{\rho}(p_2+p_3)}+
\frac{(\p_1\times\p_3)^i\p_2^j}{D_{\rho}(p_1+p_3)}+(i\leftrightarrow j)
\eeq


\end{document}